\documentclass[noeprint,twocolumn,aps,pra,showpacs,groupedaddress,superscriptaddress,nofootinbib]{revtex4-2}  
\usepackage{graphicx}   
\usepackage[dvipsnames]{xcolor} 
\usepackage{color}      
\usepackage{amsmath}    
\usepackage{amsfonts}   
\usepackage{float}      
\usepackage[pdftex,linkcolor=blue,citecolor=blue,urlcolor=blue,colorlinks]{hyperref} 
\usepackage{outlines}   

\usepackage{enumitem}
\setlist[enumerate,1]{label=(\alph*)}
\setlist[enumerate,2]{label=(\roman*)}

\usepackage{tikz}
\usetikzlibrary{calc}
\usetikzlibrary {fadings,patterns}
\usetikzlibrary{arrows.meta}  
\usetikzlibrary{decorations,calligraphy}  	
\tikzfading[name=fade1, left color = transparent!0, right color = transparent!60]
\tikzfading[name=fade2, left color = transparent!60, right color = transparent!80]

\graphicspath{{FiguresPaper/}}         
\newcommand\varpm{\mathbin{\vcenter{\hbox{%
				\oalign{\hfil$\scriptstyle+$\hfil\cr
					\noalign{\kern-.3ex}
					$\scriptscriptstyle({-})$\cr}%
}}}}
\newcommand\varmp{\mathbin{\vcenter{\hbox{%
				\oalign{$\scriptstyle({+})$\cr
					\noalign{\kern-.3ex}
					\hfil$\scriptscriptstyle-$\hfil\cr}%
}}}}

\begin{document}
	
\title{Many-body Aharonov-Bohm caging in a lattice of rings\\  
}

\author{Eul\`{a}lia Nicolau}
\affiliation{%
	Departament de F\'{i}sica, Universitat Aut\`{o}noma de Barcelona, E-08193 Bellaterra, Spain.
} 

\author{Anselmo M. Marques} 
\affiliation{Department of Physics and I3N, University of Aveiro, 3810-193 Aveiro, Portugal.} 

\author{Ricardo G. Dias} 
\affiliation{Department of Physics and I3N, University of Aveiro, 3810-193 Aveiro, Portugal.} 

\author{Jordi Mompart} 
\affiliation{Departament de F\'{i}sica, Universitat Aut\`{o}noma de Barcelona, E-08193 Bellaterra, Spain.
} 

\author{Ver\`{o}nica Ahufinger} 
\affiliation{%
	Departament de F\'{i}sica, Universitat Aut\`{o}noma de Barcelona, E-08193 Bellaterra, Spain.
}

\begin{abstract}
We study a system of a few ultracold bosons loaded into the states with orbital angular momentum $l=1$ of a one-dimensional staggered lattice of rings. Local eigenstates with winding numbers $+l$ and $-l$ form a Creutz ladder with a real dimension and a synthetic one. States with opposite winding numbers in adjacent rings are coupled through complex tunnelings, which can be tuned by modifying the central angle $\phi$ of the lattice. We analyze both the single-particle case and the few boson bound-state subspaces for the regime of strong interactions using perturbation theory, showing how the geometry of the system can be engineered to produce an effective $\pi$-flux through the plaquettes. We find non-trivial topological band structures and many-body Aharonov-Bohm caging in the $N$-particle subspaces even in the presence of a dispersive single-particle spectrum. Additionally, we study the family of models where the angle $\phi$ is introduced at an arbitrary lattice periodicity $\Gamma$. For $\Gamma>2$, the $\pi$-flux becomes non-uniform, which enlarges the spatial extent of the Aharonov-Bohm caging as the number of flat bands in the spectrum increases. All the analytical results are benchmarked through exact diagonalization.  
\end{abstract}
\maketitle

\section{Introduction}\label{SecIntroduction}
Neutral particles can emulate the dynamics of electrons in the presence of magnetic fields through the engineering of artificial gauge fields \cite{Dalibard2011,Goldman2014}. In the well-known Aharonov-Bohm effect \cite{Aharonov1959,Wu1975}, a charged particle performing a closed loop on a region with a non-zero electromagnetic potential acquires not only a dynamical phase but also an additional phase known as the Aharonov-Bohm phase. For particular periodic lattice geometries, single-particle wavefunctions undergo a sharp localization due to destructive interference known as Aharonov-Bohm caging \cite{Vidal1998,Vidal2000a}. This effect arises in systems such as the $\mathcal{T}_3$ model \cite{Vidal1998,Bercioux2009,Bercioux2011} or the diamond chain \cite{Vidal2000a}, and it has been observed in several experimental platforms, such as networks of conducting wires \cite{Abilio1999,Naud2001}, ultracold atoms \cite{Shinohara2002}, and photonic lattices \cite{Mukherjee2018,Kremer2020,Jorg2020}.

Of particular interest is the role that interactions play in a system with single-particle Aharonov-Bohm caging, which has been explored in different regimes \cite{DiLiberto2019,Gligoric2019,Vidal2000a,Creffield2010,Pelegri2020}. Addition of interactions lifts the degeneracy of the single-particle flat bands, providing a mechanism for particles to avoid caging \cite{DiLiberto2019,Vidal2000a,Creffield2010}. However, in the regime of strong interactions, Aharonov-Bohm caging of two particles can be recovered for appropriately tuned magnetic fluxes through the formation of bound states \cite{Creffield2010}. 

Here, we study a one-dimensional lattice of ring potentials populated by orbital angular momentum (OAM) modes with $l=1$ and winding numbers $\nu=\pm l$. Such states give rise to complex couplings that can be engineered by modifying the geometry of the lattice \cite{Polo2016a,Pelegri2019,Pelegri2019a,Pelegri2019b,Pelegri2019c,Pelegri2020}. Thus, it is a system where synthetic fluxes arise naturally. Ring trapping potentials can be created experimentally using a variety of techniques (see \cite{Amico2021} and references therein), and OAM can be transferred by rotating a weak link \cite{Ramanathan2011,Wright2013}, by coherent transfer of angular momentum from photons to the atoms \cite{Andersen2006,Franke-Arnold2017}, or by doing a temperature quench \cite{Corman2014a}. Alternatively, such a model can be realized by exciting atoms to the $p$ band in a conventional optical lattice \cite{Wirth2011,Li2016,Kiely2016,Kock2016}. The local eigenstates with winding number $\nu=\pm l$ provide the system with a synthetic dimension, such that it can be mapped to a Creutz ladder model with a flux threading each plaquette. For this family of models, interaction induced effects have been studied for repulsive \cite{Takayoshi2013,Tovmasyan2013,Zurita2020} and attractive \cite{Tovmasyan2016,Tovmasyan2018} on-site interactions, and for nearest-neighbor interactions \cite{,Sticlet2014,Junemann2017,Kuno2020b}. In particular, two-body Aharonov-Bohm caging was explored in \cite{Zurita2020}, where a photonic lattice implementation was proposed. Here, we explore the $N$-boson case and further generalize the study to the case of non-uniform fluxes, which are known to enrich the Aharonov-Bohm caging phenomenology in single-particle diamond lattices \cite{Mukherjee2020}.

The article is organized as follows. We introduce the system in Section \ref{SecPhysicalSystem} and analyze the single-particle case in Sec. \ref{SecSingleParticle}. For the case in which a $\pi$-flux threads each plaquette, we analyze both the topology of the system and study the Aharonov-Bohm caging effect in terms of the compact localized states (CLSs) that compose the flat-band spectrum. In Section \ref{SecNParticle}, we generalize this study to the case of $N$ particles by introducing on-site repulsive interactions and studying the regime of strong interactions using perturbation theory. In Sec. \ref{SecStaggered}, we generalize the study to the case of non-uniform fluxes and summarize our conclusions in Sec. \ref{SecConclusions}.

\section{Physical system}\label{SecPhysicalSystem}  
\begin{figure}[t]
	\includegraphics{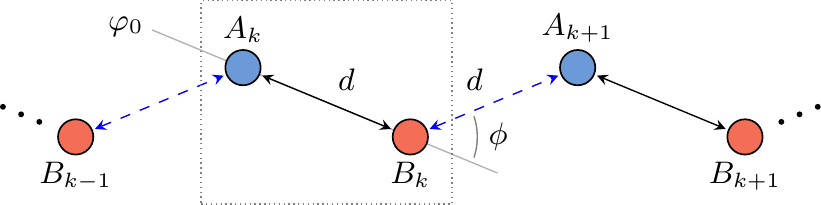}
	\caption{Diagram of the one-dimensional staggered chain where the adjacent sites $A$ and $B$ are separated by a distance $d$. The unit cell is marked by a rectangle and the grey line indicates the origin of the phase $\varphi_0$. The black arrows denote real tunneling amplitudes while the blue ones indicate complex tunneling amplitudes between states of different winding number.}\label{FigPhysicalSystem}
\end{figure}

We consider a few bosons loaded into a one-dimensional lattice  where the adjacent sites are equally separated by a distance $d$. Each unit cell $k$ is composed of two sites $A_k$ and $B_k$, and we make the lattice staggered by introducing an angle $\phi$ as depicted in Fig. \ref{FigPhysicalSystem}. Given the local polar coordinates of each site, $(\rho_{j_k},\varphi_{j_k})$ with $j=A,B$, the local trapping potential is a ring potential of the form $V(\rho_{j_k})=\frac{1}{2} M \omega^{2}(\rho_{j_k}-\rho_0)^{2}$, where $\omega$ is the frequency of the radial potential, $M$ is the mass of the particles, and $\rho_0$ is the radius. For $\rho_0=0$, the ring trap reduces to a harmonic potential and we consider identical local potentials at each site. 

The eigenstates of each isolated ring have a well-defined orbital angular momentum (OAM) $l$ with winding numbers $\nu=\pm l$. We will denote the local eigenstates as $|j_k^\nu\rangle$, where $k$ is the unit cell index,  $j=A,B$ is the site, and $\nu$ is the winding number. These sets of local eigenstates with different OAM $l$ are well-separated in energy, which makes them effectively decoupled in a lattice structure \cite{Polo2016a,Pelegri2019}. Then, the total field operator for the states with OAM $l$ in the lattice reads
\begin{equation}\label{EqWavefunctionL}
\begin{aligned}
\hat{\Psi}_l=& \sum_{k=1}^{N_{c}} \sum_{\nu=\pm l} \phi^{\nu}_{A_{k}}\left(\rho_{A_{k}}, \varphi_{A_{k}}\right) \hat{a}^{\nu}_{k}+\phi^{\nu}_{B_{k}}\left(\rho_{B_{k}}, \varphi_{B_{k}}\right) \hat{b}^{\nu}_{k}, 
\end{aligned}
\end{equation}
where $N_c$ is the number of unit cells, and $\hat{a}^{\nu}_{k}$ and $\hat{b}^{\nu}_{k}$ are the annihilation operators of the local eigenstates $|A_k^{\nu}\rangle$ and $|B_k^{\nu}\rangle$, respectively. The wavefunctions of each state $|j_k^\nu\rangle$ are given by
\begin{equation}
	\phi^{\nu}_{j_{k}}\left(\rho_{j_{k}}, \varphi_{j_{k}}\right)=\left\langle\mathbf{r} \mid j_{k}^ \nu\right\rangle=\psi\left(\rho_{j_{k}}\right) e^{i\nu\left(\varphi_{j_{k}}-\varphi_{0}\right)},
\end{equation}
where $\psi\left(\rho_{j_{k}}\right)$ is the radial part of the wavefunction and $e^{i\nu\left(\varphi_{j_{k}}-\varphi_{0}\right)}$ is the complex phase due to the non-zero OAM, with $\varphi_0$ indicating the origin of the phase. 

Consider now a single unit cell, \textit{i.e.}, two rings side by side  ($j=A,B$). The single-particle Hamiltonian restricted to a fixed value of OAM reads
\begin{equation}\label{EqTotalHamiltonian}
	\hat{\mathcal{H}}_{l}^0=\int d^2r\, \hat{\Psi}_{l}^{\dagger}\left[-\frac{\hbar^{2} \nabla^{2}}{2 M}+V(\mathbf{r})\right] \hat{\Psi}_{l},
\end{equation}
where the total potential $V(\mathbf{r})$ is the sum of the truncated  potentials of each site. The tunneling amplitudes between the states $|j^\nu_k\rangle$ with OAM $l$ are given by the overlap integrals of the corresponding wavefunctions $\phi^\nu_{j}(\rho_j,\varphi_j)$ \cite{Polo2016a}, 
\begin{equation}\label{EqCouplings}
J^{\nu,\nu'}_{j,j'}=e^{i(\nu-\nu') \varphi_{0}} \int\left(\phi^{\nu}_{j}\left(\varphi_{0}=0\right)\right)^{*} \hat{\mathcal{H}}_{l}^0\, \phi^{\nu'}_{j'}\left(\varphi_{0}=0\right) d^{2} r,
\end{equation}
where $j,j'=A,B$ identify the sites, and $\nu,\nu'=\pm l$, the winding numbers. Also, we have factorized and rewritten the wavefunctions as $\phi_{j}^{\nu}=e^{-i\nu\varphi_{0}} \phi_{j}^{\nu}\left(\varphi_{0}=0\right)$. These couplings were thoroughly analyzed in \cite{Polo2016a} by studying the mirror symmetries of the system. The authors found that there are only three distinct couplings: $J_1\equiv J_{j, j}^{\nu,-\nu}$  couples the opposite winding number OAM modes within a single ring, $J_{2} \equiv J_{A,B}^{\nu,\nu}$ couples same winding number modes in adjacent rings, and $J_{3} \equiv J_{A,B}^{\nu,-\nu}$ couples opposite winding number modes in adjacent rings. The complex factor in each coupling (\ref{EqCouplings}) is determined by the origin of the phase, $\varphi_0$, through the factor $e^{i(\nu-\nu')\varphi_0}$. For two inline rings, $\varphi_0$ can always be chosen so that the complex factor vanishes. We choose the origin of the phase along the $A_k$ and $B_k$ sites of the same unit cell (see Fig.~\ref{FigPhysicalSystem}), such that the corresponding couplings are real. The inter-cell couplings between the sites $B_k$ and $A_{k+1}$ form an angle $\phi$ with respect to the origin of the phase, such that the corresponding couplings $J_3$ and $J_1$ acquire a complex phase $e^{\pm i2l\phi}$. Therefore, one can tune the complex phase of these couplings by modifying the geometry of the staggered chain, \textit{i.e.}, the angle $\phi$ (see Fig.~\ref{FigPhysicalSystem}).

\begin{figure}[t]
	\includegraphics{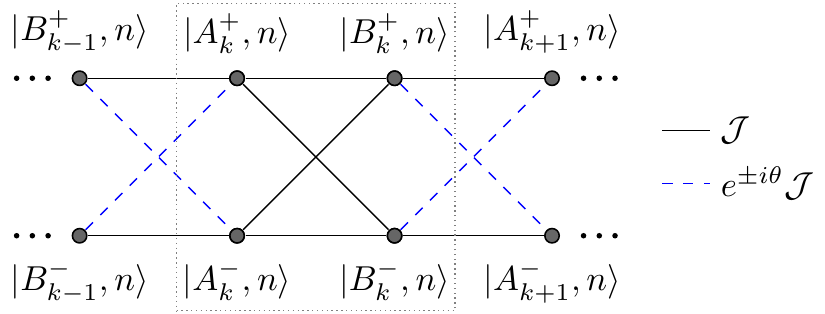}
	\caption{Schematic representation of the sites and couplings of the lattice formed by a real dimension and the synthetic dimension spanned by the two circulations $\pm $ in each site $A_k$ and $B_k$. The unit cell is indicated as a dotted rectangle and the complex couplings are $e^{i\theta}\mathcal{J}$ from circulation $+$ to $-$ and its complex conjugate in the opposite direction. }\label{FigCreutz}
\end{figure}

The couplings in a two-ring system for $l=1$ were studied in \cite{Pelegri2019}: the authors found that the magnitudes of the couplings decay with the separation distance $d$ between the two rings while the difference between $|J_3|$ and $|J_2|$ also decreases with $d$ \cite{Pelegri2019}. Additionally, $|J_1|$ is one order of magnitude smaller than $|J_2|$ and $|J_3|$ for all distances. In this work, we focus on the regime of large distances, defining $|J_2|=|J_3|\equiv J$, and we neglect the $J_1$ coupling. Also, we study the states with OAM $l=1$ and winding numbers $\nu=\pm 1$ and consider an integer number of unit cells. Henceforth, we will replace the winding number with the label of the circulation $\alpha=\pm$. Given the above assumptions and using harmonic oscillator units, the single-particle Hamiltonian of this system reads
\begin{equation}\label{EqSingleParticleBoseHubbardHamiltonian}
	\begin{aligned}
		\hat{\mathcal{H}}_{l=1}^0=\,&J\sum_{\alpha=\pm}\Bigg[ \sum_{k=1}^{N_c}\Big(\hat{a}^{\alpha \dagger}_{k} \hat{b}^{\alpha}_{k}+\hat{a}^{\alpha \dagger}_{k} \hat{b}^{-\alpha}_{k}\Big)+\\
		&\sum_{k=1}^{N_c-1}\Big(\hat{b}^{\alpha \dagger}_{k} \hat{a}^{\alpha}_{k+1}+e^{-2 \alpha i \phi} \hat{b}^{\alpha\dagger}_{k } \hat{a}^{-\alpha}_{k+1}\Big)+\mathrm{H.c.}\Bigg].
	\end{aligned}
\end{equation}
By representing the two circulations $+$ and $-$ as separate sites, one can depict this system as the Creutz ladder with vanishing vertical couplings shown in Fig. \ref{FigCreutz}. The two circulations $\alpha=\pm$ act as a synthetic dimension that constitutes the two legs of the ladder. Henceforward, we use the notation $|j_k^\alpha,n\rangle$ to denote the number of particles $n$ in the local state $|j_k^\alpha\rangle$. In the following Section, where we discuss the single-particle case, $n$ will always be $n=1$. For this case, the states in each site are $|A_k^\alpha,1\rangle$ and $|B_k^\alpha,1\rangle$, the couplings are $\mathcal{J}=J$ and $\theta=2\phi$. 

\section{Single particle}\label{SecSingleParticle} 
\begin{figure}[t]
	\includegraphics{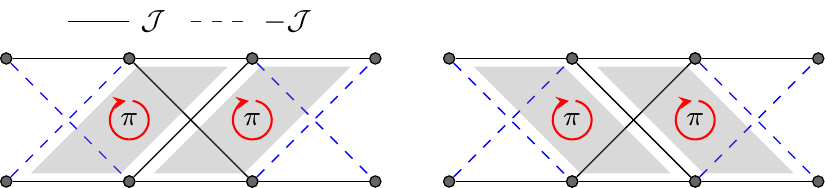}\vspace{4mm}
	\includegraphics{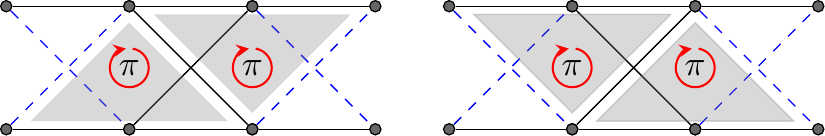}
	\caption{Schematic representation of the lattice  with a $\pi$-flux in each plaquette, for which the cross-circulation couplings reduce to $-\mathcal{J}$ (blue dashed lines). The different diagrams highlight the plaquette configurations that enclose a $\pi$-flux: rhombi and triangles with two configurations each. }\label{FigPlaquettes}
\end{figure}

\begin{figure}[h]
	\includegraphics{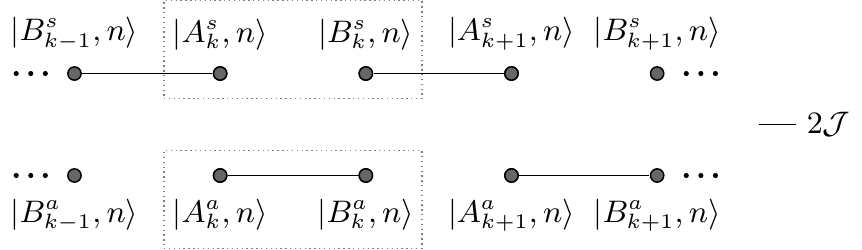}
	\caption{Decoupled symmetric and antisymmetric SSH chains with alternating couplings $2\mathcal{J}$ and $0$. The unit cell of each chain is indicated by the dotted rectangles.}
	\label{FigSSH}
\end{figure}

In this Section, we will analyze in detail the single-particle case, which will be the basis to understand the generalization to $N$ particles that we explore in Section \ref{SecNParticle}. As we have seen, the complex factor $e^{\pm 2i\phi}$ that appears in the $J_3$ couplings can be tuned by modifying the real space angle $\phi$ of the staggered chain (see Fig.~\ref{FigPhysicalSystem}). We are interested in the case $\phi=\pi/2$, for which the $J_3$ inter-cell couplings become $J_3=-J_2=-J$, thus generating a synthetic $\pi$-flux in each plaquette. Note that the couplings in the staggered chain can form either rhombus or triangle plaquettes with two configurations each, such that every one of them contains a $\pi$-flux (see Fig.~\ref{FigPlaquettes}). As a result, a particle cannot tunnel two sites to the right or to the left due to destructive interference. This destructive interference that leads to localization due to the presence of a flux is known as Aharonov-Bohm caging \cite{Vidal1998,Vidal2000a}. For $\phi=\pi/2$, the Hamiltonian in Eq.~(\ref{EqSingleParticleBoseHubbardHamiltonian}) reduces to
\begin{equation}\label{EqPiFluxHamiltonian}
	\begin{aligned}
		\hat{\mathcal{H}}^0_{l=1}=\,&J\sum_{\alpha=\pm}\Bigg[ \sum_{k=1}^{N_c}\big( \hat{a}^{\alpha \dagger}_{k} \hat{b}^{\alpha}_{k}+\hat{a}^{\alpha\dagger}_{k } \hat{b}^{-\alpha}_{k}\big)+\\
		& \sum_{k=1}^{N_c-1}\big(\hat{b}^{\alpha\dagger}_{k } \hat{a}^{\alpha}_{k+1}- \hat{b}^{\alpha\dagger}_{k } \hat{a}^{-\alpha}_{k+1}\big)+\mathrm{H.c.}\Bigg].
	\end{aligned}
\end{equation}
A topological characterization of this system can be obtained by analyzing the block-diagonalized Hamiltonian. We introduce the following basis change (with $n=1$), 
\begin{equation}\label{EqBasisChange}
	\begin{aligned}&\left|A_{k}^{s(a)},n\right\rangle=\frac{1}{\sqrt{2}}\left(\left|A_{k}^+,n\right\rangle \varpm\left|A_{k}^-,n\right\rangle\right),\\
		&\left|B_{k}^{s(a)},n\right\rangle=\frac{1}{\sqrt{2}}\left(\left|B_{k}^+,n\right\rangle \varpm\left|B_{k}^-,n\right\rangle\right),\end{aligned}
\end{equation}
that decouples the system into the two following Hamiltonians,
\begin{equation}\label{EqSSHHamiltonians}
	\begin{aligned}
		\hat{\mathcal{H}}_s=&2J \sum_{k=1}^{N_c}\hat{a}^{s\dagger}_{k } \hat{b}^{s}_{k}+\mathrm{H.c.},\\
		\hat{\mathcal{H}}_a=&2J \sum_{k=1}^{N_c-1}\hat{a}^{a\dagger}_{k+1 } \hat{b}^{a}_{k}+\mathrm{H.c.},
	\end{aligned}
\end{equation}
where $\hat{a}^{s(a)}_{k }$ and $\hat{b}^{s(a)}_{k}$ are the annihilation operators of the states in Eq. (\ref{EqBasisChange}). The Hamiltonians $\hat{\mathcal{H}}_a$ and $\hat{\mathcal{H}}_s$ correspond to two Su-Schrieffer-Heeger (SSH) chains in the dimerized limit, \textit{i.e.}, linear chains with alternating couplings where either the inter or the intra-cell coupling is zero (see Fig. \ref{FigSSH} with $n=1$ and $\mathcal{J}=J$). The two models have the same couplings, $2J$ and $0$, in opposite configurations, which leads to them having opposite topological phases. 

We consider an integer number of unit cells and that the first site of the chain is a site $A$ (and thus, the last, a site $B$), such that the edge couplings are real. In that case, the symmetric SSH chain, $\hat{\mathcal{H}}_s$, is in the trivial phase, characterized by a quantized Zak phase $\gamma=0$, and the antisymmetric chain, $\hat{\mathcal{H}}_a$, is in the topological phase with a quantized Zak phase, $\gamma=\pi$. If we instead consider a lattice starting with a $B$ site, the symmetric chain would be the one in the topological phase. Thus, for an integer number of unit cells, there are always two edge states present regardless of the configuration of the chain.

In Fig. \ref{FigCLS}(a), we represent the energy spectrum of a chain with $N_c=12$ unit cells and $\phi=\pi/2$ obtained through exact diagonalization. We obtain two flat bands and two zero-energy edge states that correspond to the superposition of the energy spectra of $\hat{\mathcal{H}}_s$ and $\hat{\mathcal{H}}_a$, in Eq.~(\ref{EqSSHHamiltonians}). The edge states are eigenstates of the antisymmetric chain and are completely localized at the edge sites (with $n=1$),
\begin{equation}\label{EqEdge}
\begin{aligned}\left|A_{1}^a,n\right\rangle_{edge}&=\frac{1}{\sqrt{2}}\left(\left|A_{1}^+,n\right\rangle -\left|A_{1}^-,n\right\rangle\right),\\
\left|B_{N_c}^a,n\right\rangle_{edge}&=\frac{1}{\sqrt{2}}\left(\left|B_{N_c}^+,n\right\rangle -\left|B_{N_c}^-,n\right\rangle\right).\end{aligned}
\end{equation}

\subsection{Single-particle Aharonov-Bohm caging}
\begin{figure}
	\includegraphics{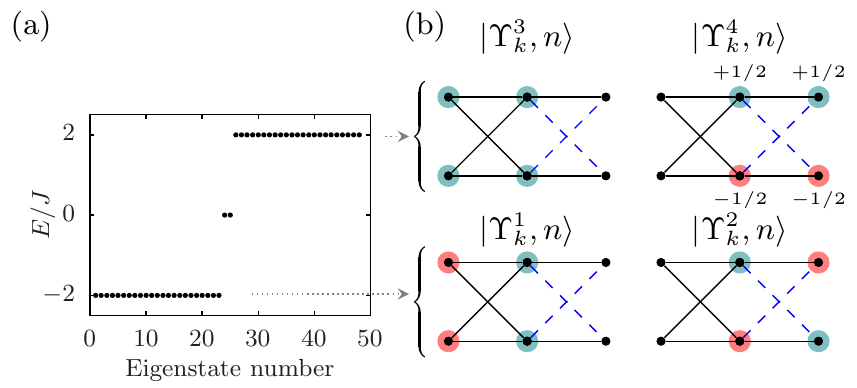}
	\caption{(a) Single-particle energy spectrum for $N_c=12$ unit cells and $\phi=\pi/2$. (b) Representation of the CLSs defined in Eq.~(\ref{EqCLS}) that are eigenstates of the Creutz ladder, see Fig.~\ref{FigCreutz}, when a $\pi$-flux threads each plaquette. The radius represents the amplitude and the color represents the phase, with red being a $\pi$ phase, and green being a phase zero.}\label{FigCLS}
\end{figure}

In this Section, we explore single-particle Aharonov-Bohm caging. The flat bands that appear in the spectrum when a $\pi$-flux threads each plaquette [see Fig.~\ref{FigCLS}(a)], are characterized by the presence of compact localized states (CLSs). These eigenstates have high real space localization: their amplitude is non-zero in a few close-by sites while being exactly zero everywhere else. The smallest possible basis for the CLSs in this model spans the states of one unit cell and an extra site (where $n=1$),
\begin{equation}\label{EqCLSbasis}
	\left\{|A_k^+,n\rangle,|A_k^-,n\rangle,|B_k^+,n\rangle,|B_k^-,n\rangle,|A_{k+1}^{+},n\rangle,|A_{k+1}^{-},n\rangle\right\}.
\end{equation}
The CLSs are found to be [see Fig.~\ref{FigCLS}(b)]
\begin{equation}\label{EqCLS}
	\begin{aligned}
		|\Upsilon_k^1,n\rangle&=\frac{1}{2}\left(|B_k^+,n\rangle+|B_k^-,n\rangle-|A_k^+,n\rangle-|A_k^-,n\rangle\right),\\
		|\Upsilon_k^2,n\rangle&=\dfrac{1}{2}\left(|B_k^+,n\rangle-|B_k^-,n\rangle-|A_{k+1}^{+}\rangle+|A_{k+1}^-,n\rangle\right),\\
		|\Upsilon_k^3,n\rangle&=\dfrac{1}{2}\left(|B_k^+,n\rangle+|B_k^-,n\rangle+|A_k^+,n\rangle+|A_k^-,n\rangle\right),\\ |\Upsilon_k^4,n\rangle&=\dfrac{1}{2}\left(|B_k^+,n\rangle-|B_k^-,n\rangle+|A_{k+1}^+,n\rangle-|A_{k+1}^-,n\rangle\right),
	\end{aligned}
\end{equation}
and their corresponding energies are $E_1=E_2=-2\mathcal{J}$ and $E_3=E_4=2\mathcal{J}$ (where $\mathcal{J}=J$ in the single-particle case). Any initial state that can be written as a superposition of these states will remain localized in the caging cell defined in (\ref{EqCLSbasis}). 

\begin{figure}[t]
	\includegraphics[width=1\columnwidth]{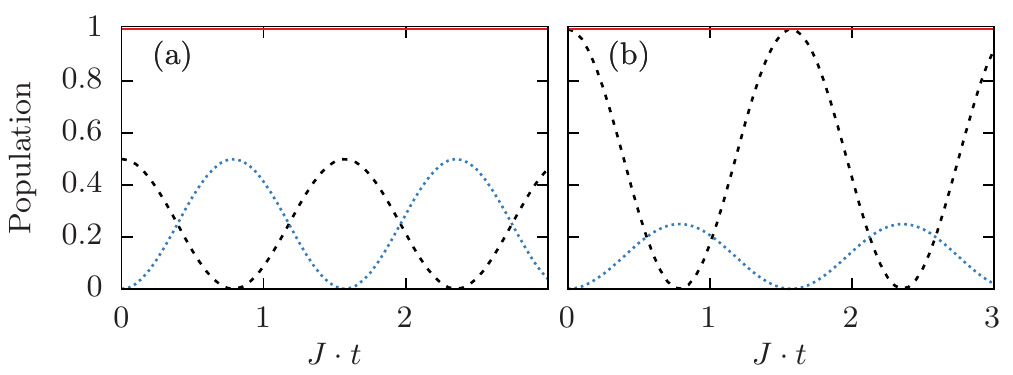}
	\caption{Time evolution of the population of the states $|j_k^\alpha,1\rangle$ with $j=A,B$ and total caged population, obtained through exact diagonalization for $J=1$, $N_c=12$ unit cells and $\phi=\pi/2$. The continuous red line is the total caged population $P_{cag}$; the dashed black line is the population in the states $|A_4^\alpha,1\rangle$, with $\alpha=\pm$; and the dotted blue line is the population in the states (a) $|B_{4}^\alpha,1\rangle$, (b)  $|B_{3}^\alpha,1\rangle$ and $|B_{4}^\alpha,1\rangle$. The initial states are (a) $\left(\left|A_{4}^+,1\right\rangle +\left|A_{4}^-,1\right\rangle\right)/\sqrt{2}$ and (b) $\left|A_{4}^+,1\right\rangle$.}\label{FigCaging1}
\end{figure}

We consider an initial state where only a single site $A_k$ in the bulk of the chain is populated.  Fig. \ref{FigCaging1}(a) shows the time evolution of the population of each local eigenstate, $P_{|j_k^{\alpha},1\rangle}$ (with $j=A,B$), for the initial state $\left(\left|A_{k}^+,1\right\rangle +\left|A_{k}^-,1\right\rangle\right)/\sqrt{2}$, which corresponds to the superposition $(|\Upsilon_k^3,1\rangle-|\Upsilon_k^1,1\rangle)/\sqrt{2}$. The population coherently oscillates between the sites $A_{k}$ and $B_{k}$ without populating any other sites due to destructive interference at $B_{k-1}$ and $A_{k+1}$. Thus, the total caged population, $P_{cag}=P_{|A_k^{+},1\rangle}+P_{|A_k^{-},1\rangle}+P_{|B_k^{+},1\rangle}+P_{|B_k^{-},1\rangle}$, stays at $P_{cag}=1$ throughout the time evolution. Additionally, the two circulations within each site maintain the same population at all times: $P_{|A_k^+,1\rangle}=P_{|A_k^-,1\rangle}$ and $P_{|B_{k}^+,1\rangle}=P_{|B_k^-,1\rangle}$. For the initial state $\left(\left|A_{k}^+,1\right\rangle -\left|A_{k}^-,1\right\rangle\right)/\sqrt{2}=(|\Upsilon_k^4,1\rangle-|\Upsilon_k^2,1\rangle)/\sqrt{2}$, one obtains identical dynamics but the exchange in population takes place between the sites $A_{k}$ and $B_{k-1}$, as the sign of the superposition shifts the destructive interference to the sites $B_k$ and $A_{k-1}$. Fig. \ref{FigCaging1}(b) shows the time evolution for the initial state $\left|A_{k}^+,1\right\rangle=(-|\Upsilon_k^1,1\rangle+|\Upsilon_k^3,1\rangle-|\Upsilon_{k-1}^2,1\rangle+|\Upsilon_{k-1}^4,1\rangle)/2$. As this initial state cannot be written as a superposition of CLSs of a single caging cell, the population reaches both the sites $B_k$ and $B_{k-1}$. The total caged population, which in this case also stays constant, is $P_{cag}=P_{|A_k^{+},1\rangle}+P_{|A_k^{-},1\rangle}+P_{|B_k^{+},1\rangle}+P_{|B_k^{-},1\rangle}+P_{|B_{k-1}^{+},1\rangle}+P_{|B_{k-1}^{-},1\rangle}$. 
Also, we simulate a chain with $N_c=12$ unit cells and choose the unit cell $k=4$ for the initial state. The caging dynamics in Fig.~\ref{FigCaging1} can also be understood in terms of the decoupled dimers of the SSH chains. For the symmetric and antisymmetric initial states, in Eq.~(\ref{EqBasisChange}), the population remains trapped in the corresponding dimer of the symmetric, $\hat{\mathcal{H}}_s$, or the antisymmetric, $\hat{\mathcal{H}}_a$, chain (see Fig.~\ref{FigSSH}). In contrast, the initial state $\left|A_{k}^+,1\right\rangle$ populates both the symmetric and antisymmetric SSH chains, such that the population reaches both dimers and as a consequence reaches a broader spatial extent.  

\section{N particle}\label{SecNParticle} 
In this Section, we explore the many-body dynamics of the system for $N$ bosons with repulsive interactions. For an ultracold and dilute gas of atoms, two-body collisions dominate, and the interaction Hamiltonian for a lattice of rings restricted to a single OAM manifold can be written as 
\begin{equation}\label{EqInteractionHamiltonian}
\hat{\mathcal{H}}^{int}_{l}=\frac{g}{2} \int d^2r\, \hat{\Psi}_{l}^{\dagger} \hat{\Psi}_{l}^{\dagger} \hat{\Psi}_{l} \hat{\Psi}_{l},
\end{equation}
where $g$ is proportional to the $s$-wave scattering length and fulfills $g>0$. Introducing the expression of the bosonic field operator, Eq.~(\ref{EqWavefunctionL}), and considering only on-site interactions, the interaction Hamiltonian for $l=1$ becomes
\begin{equation}\label{EqInteractionHamiltonianHubbard}
	\hat{\mathcal{H}}^{int}_{l=1}\hspace{-0.5mm}=\hspace{-0.5mm}\dfrac{U}{2}\hspace{-1mm}\sum_{j=A,B}\sum_{k=1}^{N_c}\!\left[ \hat{n}_{j_k}^+(\hat{n}^+_{j_k}\!-\!1)\!+\!\hat{n}^-_{j_k}(\hat{n}^-_{j_k}\!-\! 1)\!+\!4\hat{n}^+_{j_k}\hat{n}^-_{j_k}\right]\!,
\end{equation}
where $\hat{n}^{\alpha}_{j_k}=\hat{j}^{\alpha\dagger}_{k}\hat{j}^{\alpha}_{k}$ is the number operator and the interaction strength is defined as $U \equiv g \int d^2r\left|\psi\left(\rho_{j_{k}}\right)\right|^{4}$ \cite{Pelegri2019}. Besides the common Bose-Hubbard interaction terms for each of the circulations, $\alpha=\pm$, a cross-circulation term appears. Thus, this realization of a Creutz ladder yields a nearest-neighbor interaction term along the rungs of the ladder that is not usually present in other realizations of this model.

Henceforward, we will analyze the regime of strong interactions, in which the interaction term dominates over the tunneling term, $U\gg J$. We are interested in the bound-states where the $N$ bosons occupy a single site of the lattice,  $\left\{|j_k^{\alpha},n\rangle\otimes|j_k^{-\alpha},m\rangle\right\}$, where there are $n$ particles in one circulation and $m$ particles in the other circulation (with $n+m=N$). In the regime of strong interactions, the kinetic Hamiltonian, $\hat{\mathcal{H}}_{l=1}^0$ [Eq.~(\ref{EqSingleParticleBoseHubbardHamiltonian})], is introduced as a perturbation that couples the bound states $\left\{|j_k^{\alpha},n\rangle\otimes|j_k^{-\alpha},m\rangle\right\}$ in adjacent sites. This effect creates subspaces that are well-separated in energy, and thus, effectively uncoupled. We will analyze in detail the two and three-particle cases as an example in the next subsections. The matrix elements of the effective Hamiltonian of each subspace up to third order are given by \cite{Bir1974,Tannoudji1992}
\begin{equation}\label{EqEffectiveHamiltonian}
	\begin{aligned}\langle d|&\hat{\mathcal{H}}_{\mathrm{eff}}| d^{\prime}\rangle= E_{d}^{0} \delta_{d d^{\prime}}+\frac{1}{2} \sum_{w}\langle d|\hat{\mathcal{H}}_{l=1}^0| w\rangle\langle w|\hat{\mathcal{H}}_{l=1}^0| d^{\prime}\rangle\cdot\\
		&\cdot\!\!\left[\frac{1}{E_{d}^{0}-E_{w}^{0}}+\frac{1}{E_{d^{\prime}}^{0}-E_{w}^{0}}\right]+
		\\ &+\frac{1}{2} \sum_{w w^{\prime}}\langle d|\hat{\mathcal{H}}_{l=1}^0| w\rangle\langle w|\hat{\mathcal{H}}_{l=1}^0| w^{\prime}\rangle\langle w^{\prime}|\hat{\mathcal{H}}_{l=1}^0| d^{\prime}\rangle\cdot\\
		&\cdot\!\!\left[\frac{1}{\left(E_{d}^{0}-E_{w}^{0}\right)\left(E_{d}^{0}-E_{w^{\prime}}^{0}\right)}+\frac{1}{\left(E_{d^{\prime}}^{0}-E_{w}^{0}\right)\left(E_{d^{\prime}}^{0}-E_{w^{\prime}}^{0}\right)}\right]\!, \end{aligned}
\end{equation}
where $|d\rangle,|d'\rangle$ are the bound-states, $|w\rangle,|w'\rangle$ are the mediating states in each hopping process, and $E^0$ are the unperturbed energies. Note that the first-order corrections are always zero. For $|d\rangle\neq|d'\rangle$, one obtains an effective tunneling term, while for $|d\rangle=|d'\rangle$, one obtains an effective on-site potential. While Eq.~(\ref{EqEffectiveHamiltonian}) provides a good description up to $N=3$, for $N>3$, one would need to compute the higher-order terms of the perturbative expansion.

\subsection{Two and three particles}\label{SecTwoParticle} 
For the two and three-particle cases, there are only two subspaces available that arise from the following bound-state classes:
\begin{enumerate}
	\item $\mathcal{A}$: $N$ particles occupy the same site and the same circulation, $|j_k^{\alpha},N\rangle$. These are the bound-states that minimize the interaction energy, which is $E_{\mathcal{A}}=N(N-1)U/2$.
	\item $\mathcal{B}$: these bound-states maximize the interaction energy and take the following two forms:
	\begin{enumerate}
		\item For $N$ even, $N/2$ particles in each circulation, $$\left\{|j_k^{+},N/2\rangle\otimes|j_k^-,N/2\rangle\right\},$$
		with energy $E_{\mathcal{B},\rm{even}}=(3N^2/2-N)U/2$.
		\item For $N$ odd, $(N-1)/2$ particles in one circulation and $(N-1)/2+1$ in the other
		$$\hspace{12mm}\left\{
		\begin{array}{c}
			|j_k^{+},(N-1)/2\rangle\otimes|j_k^-,(N-1)/2+1\rangle,\\
			|j_k^{+},(N-1)/2+1\rangle\otimes|j_k^-,(N-1)/2\rangle
		\end{array}\right\},$$
		with a slightly lower energy, $E_{\mathcal{B},\rm{odd}}=(3N^2/2-N-1/2)U/2$.
	\end{enumerate}
\end{enumerate}

\subsubsection{$\mathcal{A}$ subspace}
We introduce the coupling $J$ as a perturbation, \textit{i.e.}, $U\gg J$, such that the states of the $\mathcal{A}$ subspace in adjacent sites become coupled. The states for the two-particle case, \textit{e.g.}  $|A_k^{\alpha},2\rangle$ and $|{B}_k^{\alpha'},2\rangle$, become coupled through second-order hopping processes, while the states in the three-particle case, \textit{e.g.}  $|A_k^{\alpha},3\rangle$ and $|{B}_k^{\alpha'},3\rangle$, become coupled through third-order hopping processes. Additionally, each state is coupled to itself also through second-order hoppings, such that an effective on-site potential arises.  Note that for both cases, the third-order contribution to the effective on-site potential is zero. Also, the on-site potential has different magnitudes for the bulk, $V_B$, and the edge, $V_E$, since the number of available mediating states for the bulk states is twice the number of the ones available for the states localized at the edge sites \cite{Bello2016,DiLiberto2016,Marques2017}. Using Eq.~(\ref{EqEffectiveHamiltonian}) up to second order for the two-particle case and up to third order for the three-particle case, the resulting effective chains become a Creutz ladder, depicted in Fig.~(\ref{FigCreutz}) with $n=2$ or $3$. The parameters that characterize the two and three-particle effective models as well as those of the single-particle case are given in Table~\ref{Table}.

\begin{table}[b]
	\begin{center}
		\begin{tabular}{c|c|c|c|c}
			& \,Single-particle\, & $\mathcal{A}_2$ & $\mathcal{A}_3$ & $\mathcal{B}_3$ \\ \hline\hline
			\rule{0pt}{12pt}
			$\mathcal{J}$ & $J$ & $2J^2/U$ & $3J^3/(2U^2)$ & $121J^3/(72U^2)$ \\[2pt] \hline 
			$\theta$ & $2\phi$ & $4\phi$ & $6\phi$ & $2\phi$ \\[2pt] \hline 
			$\phi$ & $\pi/2$ & $\pi/4$ & $\pi/2,\pi/6$ & $\pi/2$\\[2pt] \hline 
			$V_E$ & --- & $4J^2/U$ & $3J^2/U$ & $11J^2/(6U)$ \\[2pt] \hline 
			$V_B$ & --- & $8J^2/U$ & $6J^2/U$ & $11J^2/(3U)$ \\[2pt] \hline 
			$V$ & --- & $2J^2/U$ & $J^2/U$ & ---
		\end{tabular}
	\end{center}
	\caption{Summary of parameters that characterize the single-particle case and the two and three-particle effective subspaces that exhibit Aharonov-Bohm caging. Parameters of the Creutz ladder defined in Fig. \ref{FigCreutz}: couplings $\mathcal{J}$, angle $\theta$ and real space angle $\phi$ that induces a $\pi$-flux. Effective on-site potential up to second-order corrections at the edge sites, $V_E$, and the bulk sites, $V_B$, and edge correction potential $V$.}
	\label{Table}
\end{table}

The inter-cell cross couplings between the $\mathcal{A}$ subspace states with opposite circulations contain a complex factor $e^{\pm i\theta}$ (see Table~\ref{Table}). Then, for two (three) particles and the real space angle $\phi=\pi/4$ ($\phi=\pi/2$ or $\pi/6$) (see Fig~\ref{FigPhysicalSystem}), the complex factor becomes a $\pi$ phase and the effective chain acquires a $\pi$-flux in each plaquette of the Creutz ladder, see Fig.~\ref{FigPlaquettes}. Due to the similarities between the single-particle model and the effective $\mathcal{A}$ subspace, we can apply the basis-change employed for the single-particle case, taking $n=2$ or $3$ in Eq.~(\ref{EqBasisChange}). As expected, one obtains two dimerized SSH-like decoupled systems with renormalized couplings [Fig.~\ref{FigSSH} with $n=2$ or $3$ and $\mathcal{J}=2J^2/U$ or $3J^3/(2U^2)$], with additional on-site potentials inherited from the Creutz ladder, $V_B$ and $V_E$. 

\begin{figure}[t]
	\includegraphics[width=1\linewidth]{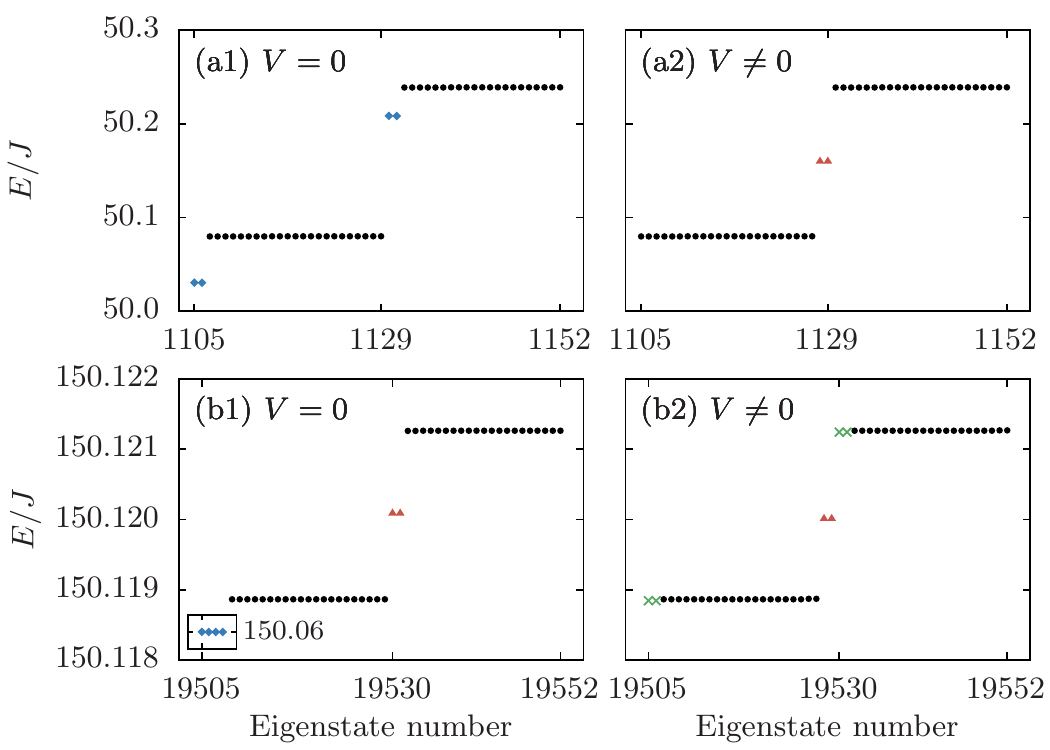}
	\caption{Energy spectrum of the $\mathcal{A}$ subspace for (a) two ($\phi=\pi/4$) and (b) three ($\phi=\pi/2$) particles, $U/J=50$ and $N_c=12$ unit cells with or without an on-site potential correction $V$ at the edge sites: (a1), (b1) $V=0$, (a2) $V=2J^2/U$, and (b2) $V=J^2/U$. We depict bulk states with black circles, Tamm-Shockley states with blue rhombi, topologically protected edge states with red triangles, and the green crosses indicate states slightly below the bulk bands.}\label{FigSpectrum2}
\end{figure}

Fig.~\ref{FigSpectrum2} shows the energy spectrum of the $\mathcal{A}$ subspace for (a1) two particles and (b1) three particles for $U/J=50$ and $N_c=12$ unit cells. We choose the angle $\phi$ that induces a $\pi$-flux in each effective Hamiltonian, $\phi=\pi/4$ and $\phi=\pi/2$, respectively. In contrast with a regular SSH model, the effective chains are not chirally symmetric due to the presence of the bulk-edge on-site potential mismatch. Therefore, the four eigenstates  that fall outside the bulk bands (blue rhombi) are non-topological Tamm-Shockley edge states, \textit{i.e.}, states induced by interactions that are localized at the edge sites due to the bulk-edge on-site potential mismatch \cite{DiLiberto2016,Bello2016,Gorlach2017,Salerno2018}. One can recover chiral symmetry in the effective model by introducing an on-site potential $V$ at the edge sites of the real space chain that exactly compensates the potential mismatch \cite{Bello2016}.  Figures~\ref{FigSpectrum2}(a2) and (b2) show the two and three-particle spectra of the $\mathcal{A}$ subspace when we introduce the on-site potential correction at the edge sites, $V=2J^2/U$ and $V=J^2/U$, respectively. In this case, we recover the spectrum of an SSH model with two symmetry-protected edge states (red triangles).

There are some differences between the two and three-particles cases. For three particles, the processes that induce the bulk-edge on-site potential mismatch are one order of magnitude higher than the ones that generate the bulk bands. Thus, the bulk-edge mismatch effectively uncouples the edge sites from the rest of the lattice, which retains chiral symmetry. Given that the symmetric and antisymmetric SSH chains are in opposite topological phases, removing the edge sites from the lattice exchanges the topological phase between the two chains. Therefore, the spectrum in Fig.~\ref{FigSpectrum2}(b1) presents not only the four Tamm-Shockley edge states (blue rhombi), well-separated energetically from the bulk bands, but also two topologically protected edge states (red triangles). When we introduce the potential correction $V=J^2/U$ in Fig.~\ref{FigSpectrum2}(b2), we exchange  the topological phases of the symmetric and antisymmetric chains. The Tamm-Shockley states are absorbed by the bulk and two topologically protected edge states remain. We can also observe two states in each band (green crosses) with slightly lower energies than the others due to fourth-order corrections to the on-site potential. These corrections are not observable in the two-particle case, see Fig.~\ref{FigSpectrum2}(a2), as the fourth-order corrections are two orders of magnitude smaller than the couplings that generate the bulk bands.

\begin{figure}[t]
	\includegraphics[width=1\columnwidth]{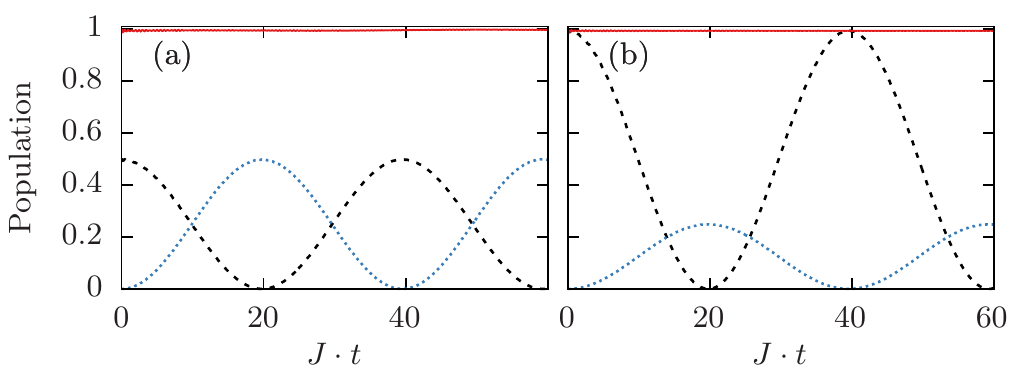}
	\includegraphics[width=1\columnwidth]{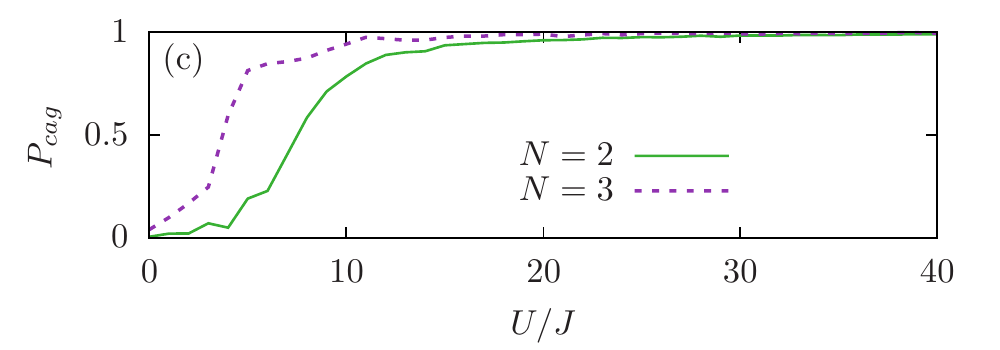}
	\caption{(a) and (b) Time evolution of the population of the states $|j_k^\alpha,2\rangle$ with $j=A,B$ and total caged population, obtained through exact diagonalization for $U/J=50$, $N_c=12$ unit cells, and $\phi=\pi/4$. The continuous red line is the total caged population $P_{cag}$; the dashed black line is the population in the states $|A_4^\alpha,2\rangle$, with $\alpha=\pm$; and the dotted blue line is the population in the states (a) $|B_{4}^\alpha,2\rangle$, (b)  $|B_{3}^\alpha,2\rangle$ and $|B_{4}^\alpha,2\rangle$. The initial states are (a) $\left(\left|A_{4}^+,2\right\rangle +\left|A_{4}^-,2\right\rangle\right)/\sqrt{2}$ and (b) $\left|A_{4}^+,2\right\rangle$. (c) Caged population, $P_{cag}$, after a time $3JT_N$ for the $\mathcal{A}$ subspace with $N=2$ and $N=3$ as a function of the ratio $U/J$. $JT_N$ is the period of the oscillations for $U/J=100$, for the two and three-particle cases and taking $\phi$ from Table \ref{Table}. The number of unit cells is $N_{c}=10$ for $N=2$ and $N_{c}=6$ for $N=3$. 
}\label{FigCaging2}
\end{figure}

Following the analogy with the single-particle case, the eigenstates of the flat-band spectra obtained for two and three particles are the CLSs in Eq.~(\ref{EqCLS}) taking $n=2$ or $3$, with energies $\pm 2\mathcal{J}$. Fig. \ref{FigCaging2} shows the time evolution of the population of the two-particle bound-states of the $\mathcal{A}$ subspace for different initial states. In particular, we consider the initial states analogous to the ones used in the single-particle case: in Fig. \ref{FigCaging2}(a), $\left(\left|A_{k}^+,2\right\rangle +\left|A_{k}^-,2\right\rangle\right)/\sqrt{2}$, and in Fig. \ref{FigCaging2}(b), $\left|A_{k}^+,2\right\rangle$. One can see that the dynamical evolution is identical to the one observed for a single particle (see Fig. \ref{FigCaging1}). In this case, the dynamics correspond to two-particle Aharonov-Bohm caging and they take place over a much longer timescale. This is because the couplings of the effective Creutz ladder are a second-order effect and, thus, much smaller in magnitude than the ones in the single-particle case (see Table~\ref{Table}). We define the total caged population as the sum of the population in a series of states: (a) $P_{cag}=P_{|A_k^{+},2\rangle}+P_{|A_k^{-},2\rangle}+P_{|B_k^{+},2\rangle}+P_{|B_k^{-},2\rangle}$; (b) $P_{cag}=P_{|A_k^{+},2\rangle}+P_{|A_k^{-},2\rangle}+P_{|B_k^{+},2\rangle}+P_{|B_k^{-},2\rangle}+P_{|B_{k-1}^{+},2\rangle}+P_{|B_{k-1}^{-},2\rangle}$. The total caged population reveals slight population losses that are due to higher-order corrections to the effective model that make the flat bands in Fig.~\ref{FigSpectrum2} slightly dispersive.

For three particles and the analogous initial states, $\left(\left|A_{k}^+,3\right\rangle +\left|A_{k}^-,3\right\rangle\right)/\sqrt{2}$ and $\left|A_{k}^+,3\right\rangle$, we obtain identical (albeit slower) dynamics that correspond to three-particle Aharonov-Bohm caging. The periods of the oscillations for the different numbers of particles and $U/J=50$ are $JT_{N=1}=1.55$, $JT_{N=2}=39.5$, $JT_{N=3}=2600$.

To further compare the two and three-particle Aharonov-Bohm caging, we consider an initial state in the $\mathcal{A}$ subspace,  $(|B_{k}^{+},n\rangle+|B_k^{-},n\rangle)/\sqrt{2}$ (with $n=2$ or $n=3$), located at the middle of the lattice, and we let it evolve through time. The caged population for this initial state is 	$P_{cag}=P_{|A_k^{+},n\rangle}+P_{|A_k^{-},n\rangle}+P_{|B_k^{+},n\rangle}+P_{|B_k^{-},n\rangle}$. Fig.~\ref{FigCaging2}(c) shows the caged population after a time $3JT_N$, where $JT_N$ is the period of the oscillations for $U/J=100$, as a function of the ratio $U/J$ for the two and three-particle cases. The caged population rapidly increases for $U>J$, reaching a value close to $1$ as the system enters the regime of strong interactions. The growth of the caged population is faster for the three-particle subspace compared to the two-particle case, and it saturates at a smaller value of $U/J$. This can be understood by inspecting the higher-order terms of the perturbative expansion. As the ratio $U/J$ decreases, higher-order terms of the perturbative expansion have to be taken into account. For two particles (and also for any subspace with an even number of particles), the odd-order perturbative corrections are always zero. Then, the next perturbative correction is fourth order, and it leads to effective on-site potentials, nearest-neighbor hoppings, and also next-nearest neighbor hoppings that destroy the CLSs. In contrast, the fourth-order correction to the three-particle case only induces an effective on-site potential, and the fifth order induces nearest-neighbor hopping terms that maintain the Creutz ladder structure that exhibits flat bands. It is not until the sixth-order correction, that the next-nearest neighbor hoppings appear, making the CLSs disappear. Thus, the three-particle subspaces are more resilient to deviations from the regime of strong interactions than the two-particle $\mathcal{A}$ subspace.

\subsubsection{$\mathcal{B}$ subspace}
\begin{figure}[t]
	\includegraphics[width=1\linewidth]{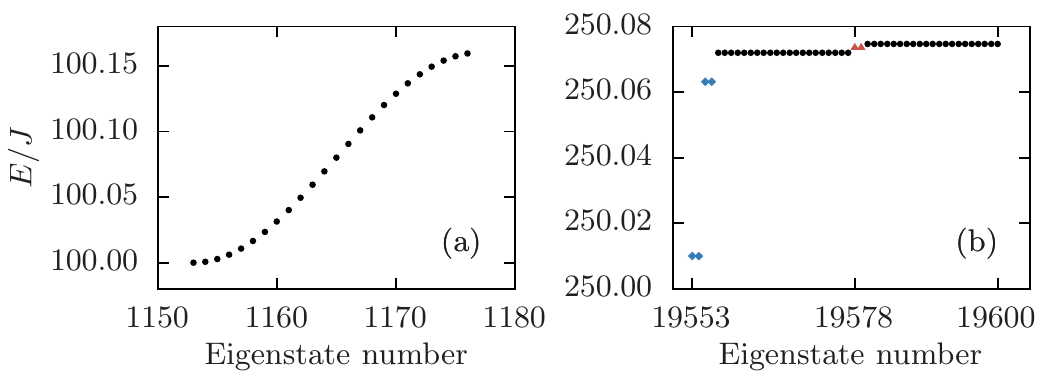}
	\caption{Energy spectrum of the  $\mathcal{B}$ subspace for (a) two ($\phi=\pi/4$) and (b) three ($\phi=\pi/2$) particles, $U/J=50$ and $N_c=12$ unit cells. We depict bulk states with black circles, Tamm-Shockley states with blue rhombi, and topologically protected edge states with red triangles.}\label{FigSpectrumB}
\end{figure}

The bound-states of the $\mathcal{B}$ subspace for the two-particles case have one particle in each circulation, $|j_k^{+},1\rangle\otimes|j_k^-,1\rangle$. When we consider the couplings between states in adjacent sites, \textit{e.g.} between $|A_k^{+},1\rangle\otimes|A_k^-,1\rangle$ and $|B_k^{+},1\rangle\otimes|B_k^-,1\rangle$, there is no complex factor, as any hopping process between opposite circulations will necessarily be followed by a hopping process with the opposite phase factor. This results in an effective linear chain with uniform couplings $2J^2/U$ and on-site potentials $V_B=4J^2/U$ at the bulk and $V_E=2J^2/U$ at the edges. Therefore, the two-particle $\mathcal{B}$ subspace has a dispersive spectrum for any $\phi$ [see Fig.~\ref{FigSpectrumB}(a)] and therefore cannot exhibit Aharonov-Bohm caging. 

The three-particle $\mathcal{B}$ subspace arises from bound states of the form $|j_k^{\alpha},2\rangle\otimes|j_k^{-\alpha},1\rangle$. In analogy with the $\mathcal{A}$ subspace cases, the $\mathcal{B}$ effective subspace is a Creutz ladder with a bulk-edge on-site potential mismatch that can be mapped to two decoupled SSH-like chains with the same on-site potential mismatch (see Table~\ref{Table}). Fig. \ref{FigSpectrumB}(b) shows the energy spectrum for the three-particle $\mathcal{B}$ subspace for $U/J=50$, $N_c=12$ unit cells, and $\phi=\pi/2$. However, in this case there is an extra ingredient: the two bound-states in the same site, $|j_k^{\alpha},2\rangle\otimes|j_k^{-\alpha},1\rangle$ and $|j_k^{\alpha},1\rangle\otimes|j_k^{-\alpha},2\rangle$, are also coupled through second-order processes that generate a complex vertical coupling in the effective Creutz model. For the angle $\phi$ that induces a $\pi$-flux, $\phi=\pi/2$, the complex couplings of each mediating process cancel with the symmetric mediating process (\textit{i.e.} inverting the direction of the hopping processes from right to left). This compensation does not occur on the edge sites, which results in an energy mismatch between the Tamm-Shockley states (blue rhombi) of the two edges. In analogy with the three-particle $\mathcal{A}$ subspace [see Fig.~\ref{FigSpectrum2}(b1)], there are two topologically protected edge states (red triangles) besides the Tamm-Shockley states.

\subsection{$N$-particle generalization}
\begin{figure}[t]
	\includegraphics{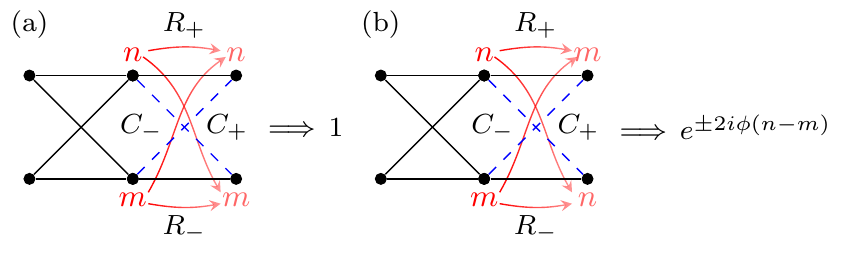}
	\caption{Hopping processes of an arbitrary $N$-particle bound state $\left\{|B_k^{\alpha},n\rangle\otimes|B_k^{-\alpha},m\rangle\right\}$ that couples to the bound-states in the adjacent site (a) $\left\{|A_{k+1}^{\alpha},n\rangle\otimes|A_{k+1}^{-\alpha},m\rangle\right\}$ and (b) $\left\{|A_{k+1}^{\alpha},m\rangle\otimes|A_{k+1}^{-\alpha},n\rangle\right\}$ and corresponding phase factors. $R_\alpha$ and $C_\alpha$ are the numbers of real and complex hopping processes, respectively, coming from each circulation and the labels $n$ and $m$ denote the number of particles in each site.  }\label{FigNhoppings}
\end{figure}

From the above cases, one can deduce a recipe to obtain Aharonov-Bohm caging in any $N$-particle subspace by looking at the $N$-particle tunneling processes involving complex tunnelings, \textit{i.e.}, the cross-circulation couplings $J_3$. We define an arbitrary bound state $\left\{|j_k^{\alpha},n\rangle\otimes|j_k^{-\alpha},m\rangle\right\}$ with $n$ particles in one circulation and $m$ particles in the other circulation such that $n+m=N$. In the regime of strong interactions, Aharonov-Bohm caging can exist in the subspace generated by these bound-states if all the $N$-particle hopping processes involving a complex phase acquire the same total phase factor, such that by appropriately choosing the angle $\phi$, one can induce a $\pi$-flux. The bound-states in the sites $B_k$ will be coupled in the adjacent sites $A_{k+1}$ (see Fig.~\ref{FigNhoppings}) through the integer number of real hoppings from each circulation, $R_\alpha$ and $R_{-\alpha}$, and the integer number of complex hoppings from each circulation, $C_\alpha$ and $C_{-\alpha}$, such that
\begin{equation}\label{EqNparticleOut}
	n=R_\alpha+C_\alpha \qquad \text{and} \qquad m=R_{-\alpha}+C_{-\alpha}.
\end{equation}
Then, the total complex factor will be given by $e^{\pm 2i\phi (C_\alpha-C_{-\alpha})}$. These states are coupled to both the bound-states $\left\{|A_{k+1}^{\alpha},n\rangle\otimes|A_{k+1}^{-\alpha},m\rangle\right\}$ [Fig.~\ref{FigNhoppings}(a)] and  $\left\{|A_{k+1}^{\alpha},m\rangle\otimes|A_{k+1}^{-\alpha},n\rangle\right\}$ [Fig.~\ref{FigNhoppings}(b)] in the adjacent site, thus fulfilling the following conditions for each case,
\begin{equation}\label{EqNparticleIn}
	\begin{aligned}
		\left\{|A_{k+1}^{\alpha},n\rangle\otimes|A_{k+1}^{-\alpha},m\rangle\right\}:&\left\lbrace
		\begin{aligned}
			n=C_{-\alpha}+R_{\alpha} \\
			m=C_{\alpha}+R_{-\alpha} 
		\end{aligned}\right\rbrace, \\ \left\{A_{k+1}^{\alpha},m\rangle\otimes|A_{k+1}^{-\alpha},n\rangle\right\}:&
	\left\lbrace
		\begin{aligned}
			n=R_{-\alpha}+C_{\alpha}\\
			m=R_{\alpha}+C_{-\alpha}
		\end{aligned}\right\rbrace.
\end{aligned}
\end{equation}
Combining Eqs.~(\ref{EqNparticleOut}) and (\ref{EqNparticleIn}), we obtain the following relations between the number of complex couplings $C_\alpha$ and the corresponding phase factors (see Fig.~\ref{FigNhoppings}),
\begin{equation}\label{EqFactorsArbitrarySubspace}
	\begin{aligned}
		\left\{|A_{k+1}^{\alpha},n\rangle\otimes|A_{k+1}^{-\alpha},m\rangle\right\}: C_{\alpha}&=C_{-\alpha}  \quad\Longrightarrow\quad 1, \\
	    \left\{|A_{k+1}^{\alpha},m\rangle\otimes|A_{k+1}^{-\alpha},n\rangle\right\}: C_{\alpha}&-C_{-\alpha}=n-m  & \\
	    &\quad\Longrightarrow\quad  e^{\pm 2i\phi (n-m)}.
	\end{aligned}
\end{equation}
Therefore, one can obtain an effective Creutz ladder model up to $N$-th order perturbation theory for any subspace with $n\neq m$. In this case, the states in the same site $\left\{|j_{k}^{\alpha},n\rangle\otimes|j_{k}^{-\alpha},m\rangle\right\}$ and $\left\{|j_{k}^{\alpha},m\rangle\otimes|j_{k}^{-\alpha},n\rangle\right\}$ are also coupled, which produces an effective vertical coupling in the Creutz ladder. The order of these couplings is $2|n-m|$ and they are in general complex. The effect of these couplings can be neglected if $2|n-m|\gg n+m=N$, as  $N$ is the order of the other couplings that compose the Creutz ladder. Alternatively, the vertical couplings vanish in the bulk for $\phi=\pi/2$, as each $N$-particle hopping process cancels with its left-right symmetric counterpart. Then, considering the vertical coupling and using Eq.~(\ref{EqFactorsArbitrarySubspace}), one can obtain a $\pi$-flux through the plaquettes by choosing
\begin{equation}\label{EqPhiNparticle}
	\left\lbrace
	\begin{aligned}
		\phi&=\frac{\pi}{2(n-m)}, &\quad\text{if}\quad 2|n-m|\gg n+m=N\\
		\phi&=\dfrac{\pi}{2}, &\quad\text{if}\quad n-m \text{ is odd.}
	\end{aligned}\right.
\end{equation} 
For $n=m$, there is only one type of bound state, $\left\{|j_{k+1}^{\alpha},n\rangle\otimes|j_{k+1}^{-\alpha},n\rangle\right\}$, such that the effective model is a linear chain with real  couplings, and the system cannot exhibit Aharonov-Bohm caging. For the $N$-particle subspaces that exhibit flat bands with $\phi\neq\pi/2$, the single-particle spectrum is dispersive, which makes these Aharonov-Bohm caging phenomena a many-body effect.

Let us see some examples. For the $\mathcal{A}$ subspaces, $N$ particles will accumulate a complex phase $e^{\pm 2iN\phi}$ when coupling the states $|B_k^{\alpha},N\rangle$ and $|A_{k+1}^{-\alpha},N\rangle$. For $N$ even, flat bands arise for $\phi=\pi/(2N)$, while for $N$ odd both $\phi=\pi/(2N)$ and $\phi=\pi/2$ yield a $\pi$-flux. Additionally, the vertical couplings are $2N$-order connections and thus, always negligible. For the $\mathcal{B}$ subspaces with an even number of particles, $N/2$, in each circulation, Aharonov-Bohm caging cannot occur. The complex phases accumulated by the particles cancel out such that all the couplings of the effective chain are real and the resulting energy bands are dispersive. However, for $N$ odd, the tunneling process of one of the particles is not compensated, leading to a complex factor $e^{\pm  2i\phi}$. Then, a phase $\phi=\pi/2$ leads to a flat-band spectrum while at the same time canceling the vertical couplings. For a real space angle $\phi=\pi/2$, the single-particle spectrum exhibits flat bands, and both the $N$ odd $\mathcal{A}$ and $\mathcal{B}$ subspaces also present a flat-band spectrum. However, for an angle $\phi=\pi/(2N)$ the $\mathcal{A}$ subspace presents flat bands in the absence of a single-particle flat-band spectrum, making this instance of Aharonov-Bohm caging a purely many-body effect. 

As one increases the number of particles in the system, the number of bound-state configurations increases and, in particular, other semi bound-states appear where not all particles are located in a single-site, \textit{i.e.}  $\left\{|j_k^{\alpha},n\rangle\otimes|j_k^{-\alpha},m\rangle\right\}$ with $n+m<N$ and $N-(n+m)$ particles not bound to the site $j$. The picture described above will hold as long as the subspaces induced by bound-states do not become degenerate with the subspaces induced by these semi bound-states. For the $\mathcal{B}$ subspaces, as their bound-states have the maximum possible energy, they will not become degenerate with any other subspace. The other subspaces can become degenerate with a subspace with some particles in a bound state in the same site, and some in other sites of the lattice. However, these instances are rare: up to ten particles, only $8$ out of $34$ bound-states are degenerate, for example, $\left\{|j_i^{\alpha},5\rangle\right\}$ and $\left\{|j_i^{\alpha},2\rangle\otimes|j_i^{-\alpha},2\rangle\right\}$. We have checked numerically the recipe to obtain $\pi$-fluxes in arbitrary subspaces given in Eq.~(\ref{EqPhiNparticle}) up to six particles.

\vspace{-3mm}
\section{Generalization to non-uniform fluxes}\label{SecStaggered}
\begin{figure}[t]
	\includegraphics{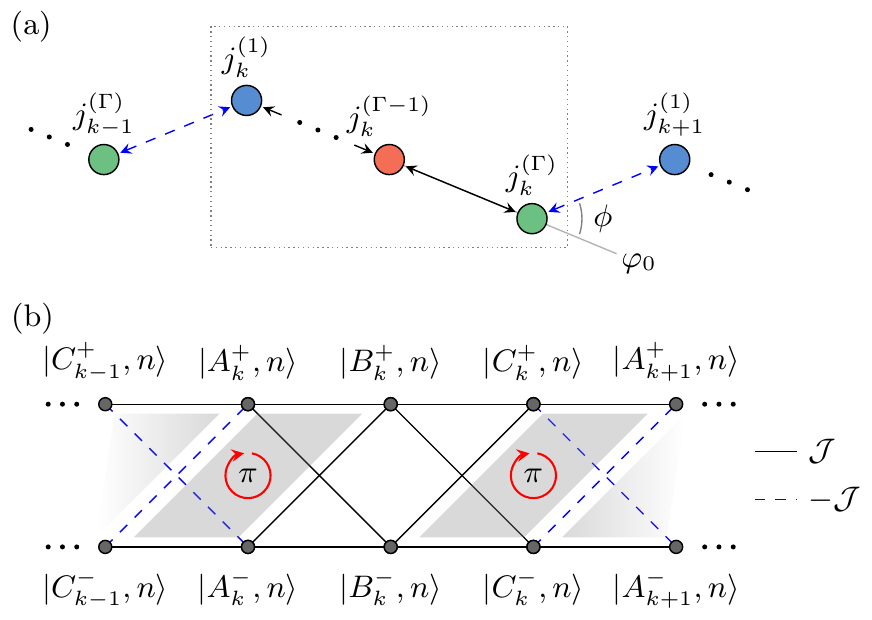}
	\caption{(a) Diagram of the one-dimensional staggered chain for an arbitrary periodicity $\Gamma$. The unit cell $k$ contains $\Gamma$ sites $\{j_k^{(1)},j_k^{(2)},...,j_k^{(\Gamma-1)},j_k^{(\Gamma)}\}$ and is enclosed by a dotted rectangle. The grey line indicates the origin of the phase $\varphi_0$ such that an angle $\phi$ is introduced in the inter-cell couplings. The black arrows denote real tunneling amplitudes while the blue ones indicate complex tunneling amplitudes between states of different winding number. (b) Schematic representation of the sites and couplings of the lattice for $\Gamma=3$ and an angle $\phi$ such that a non-uniform $\pi$-flux arises.}\label{FigSystemTau}
\end{figure}

In this Section, we generalize the study to the family of models where the angle $\phi$ of the staggered chain is introduced with an arbitrary lattice periodicity $\Gamma$, thus increasing the number of sites per unit cell [see Fig.~\ref{FigSystemTau}(a)]. The complex couplings between adjacent sites only occur between the last site of the unit cell and the first site of the next unit cell. Thus, the flux induced by this angle $\phi$ will not be present in each plaquette, with the exact flux pattern being a function of the number of sites in the unit cell. Non-uniform fluxes have been studied in diamond lattices \cite{Mukherjee2020,Li2020}, where it has been shown to lead to an enriched Aharonov-Bohm caging phenomenology. 

\begin{figure}[h]
	\includegraphics[width=1\columnwidth]{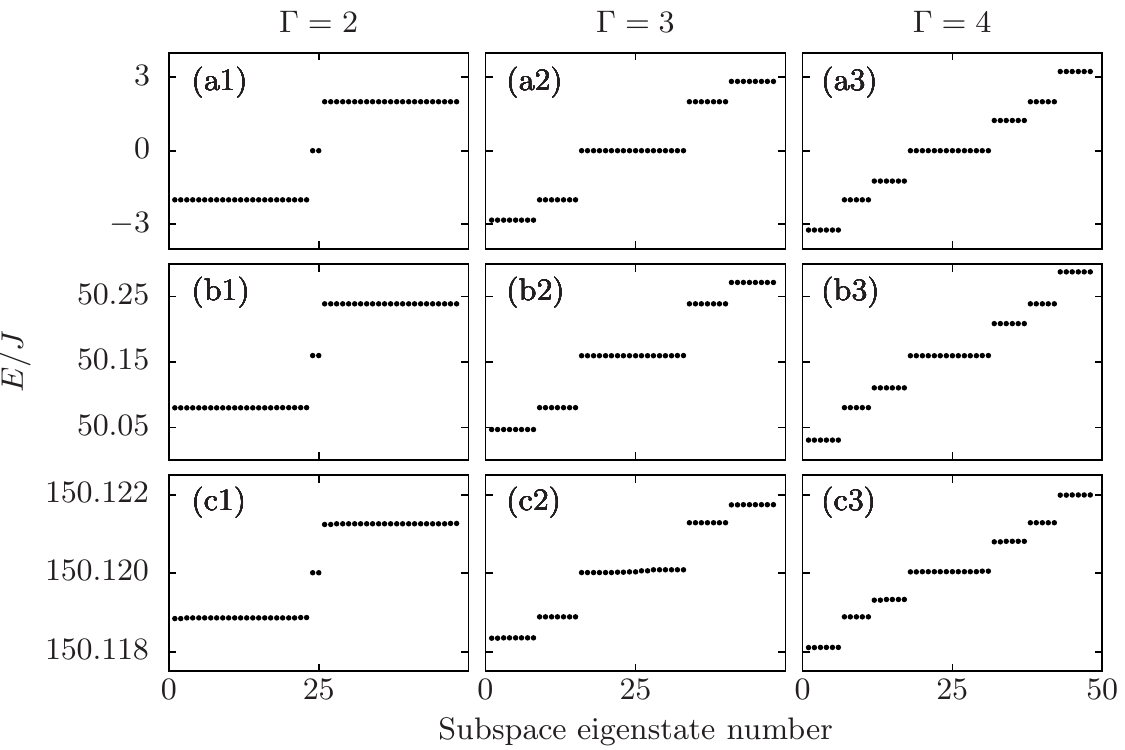}
	\caption{Energy spectrum for different number of particles (a) $N=1$, (b) $N=2$, (c) $N=3$ and periodicities (1) $\Gamma=2$, (2) $\Gamma=3$, and (3) $\Gamma=4$, for $24$ sites. For the two and three-particle cases, only the $\mathcal{A}$ subspace is shown, and we fix $U/J=50$ and introduce the on-site potential correction $V$ at the edge sites. The angle $\phi$ is taken from Eq.~(\ref{EqPhiNparticle}) such that a $\pi$-flux is obtained in each subspace: (1) $\phi=\pi/2$, (2) $\phi=\pi/4$, and (3) $\phi=\pi/2$.}\label{FigSpectrumTau}
\end{figure}

The analysis of Section \ref{SecNParticle} for the dynamics of $N$ particles in the regime of strong interactions applies also to this family of models. In particular, the angles given in Eq.~(\ref{EqPhiNparticle}) for each $N$-particle subspace also yield $\pi$-fluxes, that, in this case, are non-uniform [see an example for $\Gamma=3$ in Fig.~\ref{FigSystemTau}(b)]. The non-uniform pattern is composed of $\Gamma-2$ rhombi (or triangles) without a flux followed by two rhombi (or triangles) with a $\pi$-flux. For the case of $\Gamma=2$, discussed in Sections \ref{SecSingleParticle} and \ref{SecNParticle}, the number of rhombi plaquettes without flux is zero. As a result of the non-uniform flux pattern, a particle cannot tunnel $\Gamma$ sites to the right or the left due to destructive interference, and as a consequence, the spectrum is composed of a series of flat bands. Fig.~\ref{FigSpectrumTau} shows the energy spectrum for the single-particle case and the two and three-particle $\mathcal{A}$ subspaces for different periodicities, $\Gamma=2,3$ and $4$. The angles $\phi$, as given by Eq.~(\ref{EqPhiNparticle}), yield a $\pi$-flux, and we take $U/J=50$ and simulate $24$ sites for each case. Notably, by increasing the periodicity $\Gamma$, the number of flat bands increases, as the caging cell is enlarged and gives support to a larger number of CLSs.  The zero-energy edge states that are present for $\Gamma=2$, are buried in the central band of the spectrum for $\Gamma>2$.  As an example, we discuss the case of $\Gamma=3$ in the next subsection. 

\subsection{Example: $\Gamma=3$}
\begin{figure}[t]
	\includegraphics{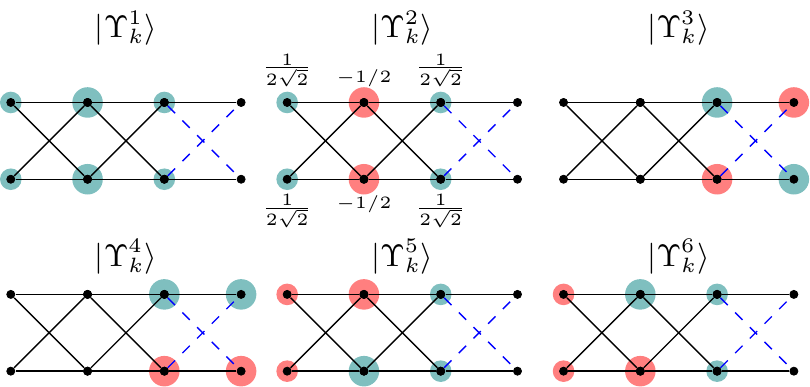}
	\caption{Representation of the CLSs for $\Gamma=3$ defined in Eq.~(\ref{EqCLStau3}) that are eigenstates of the Creutz ladder with a non-uniform $\pi$-flux, see Fig.~\ref{FigSystemTau}(b). The radius represents the amplitude and the color represents the phase, with red being a $\pi$ phase, and green being a phase zero.}\label{FigCLSTau}
\end{figure}

\begin{figure}[h]
	\includegraphics[width=1\columnwidth]{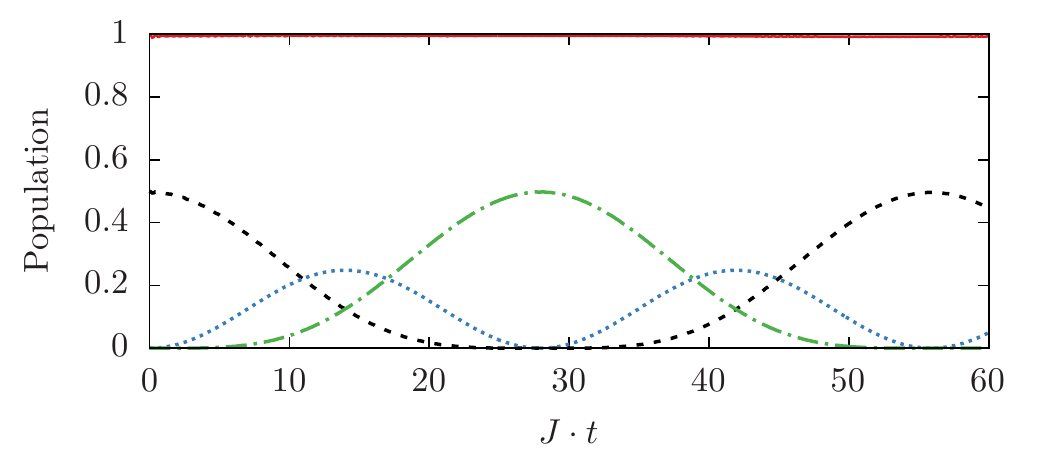}
	\caption{Time evolution of the population of the states $|j_k^\alpha,2\rangle$ with $j=A,B,C$ and total caged population $P_{cag}$ (continuous red line), obtained through exact diagonalization for $U/J=50$, $N_c=12$ unit cells and $\phi=\pi/4$. The dashed black line is the population in the states $|A_4^\alpha,2\rangle$, with $\alpha=\pm$, the dotted blue line is the population in the states $|B_{4}^\alpha,2\rangle$, and the dashed-dotted green line is the population in $|C_{4}^\alpha,2\rangle$. The initial state is $\left(\left|A_{4}^+,2\right\rangle +\left|A_{4}^-,2\right\rangle\right)/\sqrt{2}$.}\label{FigCagingTau}
\end{figure}

For a periodicity $\Gamma=3$, the unit cell has three sites that we will call $A$, $B$, and $C$. From Figures ~\ref{FigSpectrumTau}(a2), (b2), and (c2), one can see that the $N$-particle subspaces (with the appropriate $\pi$-flux inducing angle $\phi$) present six flat bands with two degenerate zero-energy bands. The eigenstates in these flat bands consist of a series of CLSs that one can find through the diagonalization of a small lattice. Analogously to the $\Gamma=2$ case, the basis states that compose the smallest caging cell are those within a unit cell and the next site

\begin{equation}\label{EqCLSbasisTau}
	\left\{\begin{aligned}
		|A_k^+,n\rangle,|A_k^-,n\rangle,|B_k^+,n\rangle,|B_k^-,n\rangle,\\
		|C_k^+,n\rangle,|C_k^-,n\rangle,|A_{k+1}^{+},n\rangle,|A_{k+1}^{-},n\rangle
	\end{aligned}\right\}.
\end{equation} 
We give below the analytical expressions of the CLSs (dropping the label $n$ for conciseness) and give a visual representation in Fig.~\ref{FigCLSTau},
\begin{equation}\label{EqCLStau3}
	\begin{aligned}|\Upsilon_k^1\rangle&=\dfrac{|A_+^k\rangle+|A_-^k\rangle+\sqrt{2}|B_+^k\rangle+\sqrt{2}|B_-^k\rangle+|C_+^k\rangle+|C_-^k\rangle}{2\sqrt{2}},\\
	|\Upsilon_k^2\rangle&=\dfrac{|A_+^k\rangle+|A_-^k\rangle-\sqrt{2}|B_+^k\rangle-\sqrt{2}|B_-^k\rangle+|C_+^k\rangle+|C_-^k\rangle}{2\sqrt{2}},\\
	|\Upsilon_k^3\rangle&=\dfrac{|C_+^k\rangle-|C_-^k\rangle-|A_+^{k+1}\rangle+|A_-^{k+1}\rangle}{2},\\
	|\Upsilon_k^4\rangle&=\dfrac{|C_+^k\rangle-|C_-^k\rangle+|A_+^{k+1}\rangle-|A_-^{k+1}\rangle}{2},\\
	|\Upsilon_k^5\rangle&=\dfrac{|C_+^k\rangle+|C_-^k\rangle-|A_+^k\rangle-|A_-^k\rangle-\sqrt{2}|B_+^k\rangle+\sqrt{2}|B_-^k\rangle}{2\sqrt{2}},\\
	|\Upsilon_k^6\rangle&=\dfrac{|C_+^k\rangle+|C_-^k\rangle-|A_+^k\rangle-|A_-^k\rangle+\sqrt{2}|B_+^k\rangle-\sqrt{2}|B_-^k\rangle}{2\sqrt{2}}.\end{aligned}
\end{equation}
The energies of the CLSs are given by 
\begin{equation}
	\begin{aligned} E_1&=2\sqrt{2}\mathcal{J}, & \quad E_2&=-2\sqrt{2}\mathcal{J},\quad & E_3&=-2\mathcal{J}, \\
		E_4&=2\mathcal{J}, & E_5&=0, & 		E_6&=0.
	\end{aligned}
\end{equation}
Let us compare these CLSs with those obtained for $\Gamma=2$, in Eq.~(\ref{EqCLS}). For $\Gamma=3$, the unit cell is enlarged, and we obtain more CLSs (six for $\Gamma=3$ vs. four for $\Gamma=2$) that also span a larger number of sites. As a direct consequence, the caging dynamics resulting from these flat bands have larger support over the lattice. To give an example, we consider the two-particle $\mathcal{A}$ subspace with $\phi=\pi/4$, $U/J=50$ and $N_c=12$ unit cells for $\Gamma=3$. In Fig.~\ref{FigCagingTau}, we show the time evolution of the population of the states, $P_{|j_k^\alpha,2\rangle}$ for the initial state $\left(\left|A_{4}^+,2\right\rangle +\left|A_{4}^-,2\right\rangle\right)/\sqrt{2}$. The red line indicates the caged population $P_{cag}=P_{|A_k^{+},2\rangle}+P_{|A_k^{-},2\rangle}+P_{|B_k^{+},2\rangle}+P_{|B_k^{-},2\rangle}+P_{|C_{k}^{+},2\rangle}+P_{|C_{k}^{-},2\rangle}$. The population oscillates between the sites $A_k$, $B_k$, $C_k$ of a single unit cell, as the  destructive interference occurs at the sites $C_{k-1}$ and $A_{k+1}$.

\section{Conclusions}\label{SecConclusions}
We have studied a system of bosons in a staggered lattice with ring traps in each site and considered the local eigenstates with orbital angular momentum $l=1$. The system can be mapped to a Creutz ladder with a real and a synthetic dimension, in which the flux enclosed in each plaquette is determined by the angle $\phi$ that makes the lattice staggered. In the single-particle case, one can tune the angle $\phi$ to obtain a uniform $\pi$-flux threading each plaquette. This leads to a flat-band spectrum characterized by the presence of CLSs and the system exhibits Aharonov-Bohm caging.

For $N$ particles in the regime of strong on-site interactions, bound-states arise where the $N$ particles populate a single site. Using perturbation theory, most of the $N$-particle subspaces can be mapped to an effective Creutz ladder with a flux that depends on the angle $\phi$. We have identified the conditions under which these subspaces present a $\pi$-flux that leads to flat bands and Aharonov-Bohm caging. Remarkably, some of these subspaces can exhibit Aharonov-Bohm caging even in the presence of a single-particle dispersive spectrum, making these instances a purely many-body effect.

Finally, we have generalized this study to the case of non-uniform fluxes by introducing the angle $\phi$ at an arbitrary lattice periodicity $\Gamma$. In this case, one can engineer flat-band spectra for different $N$-particle subspaces and an arbitrary $\Gamma$. As the unit cell increases in size, the number of flat bands increases, resulting in a larger number of CLSs that also have a greater spatial extent. As a result, the caged particles can explore a broader region of the lattice before encountering destructive interference, making the periodicity $\Gamma$ a tunable parameter that controls the spatial extent of the Aharonov-Bohm caging.

\section{Acknowledgments}
EN, VA, and JM acknowledge support through the Spanish Ministry of Science and Innovation (MINECO) (PID2020-118153GB-I00), the Catalan Government (Contract No. SGR2017-1646), and the European Union Regional Development Fund within the ERDF Operational Program of Catalunya (project QUASICAT/QuantumCat). EN acknowledges financial support from MINECO through the grant PRE2018-085815 and from COST through Action CA16221. AMM and RGD acknowledge financial support from the Portuguese Institute for Nanostructures, Nanomodelling and Nanofabrication (i3N) through Projects No. UIDB/50025/2020, No. UIDP/50025/2020, and No. LA/P/0037/2020, and funding from FCT–Portuguese Foundation for Science and Technology through Project No. PTDC/FISMAC/29291/2017. AMM acknowledges financial support from the FCT through the work Contract No. CDL-CTTRI147-ARH/2018 and from i3N through the work Contract No. CDL-CTTRI-46-SGRH/2022.

\bibliography{/Users/eulalianicolau/Dropbox/library.bib}  

\end{document}